\input harvmac
\input epsf.tex


\lref\WittenYC{
E.~Witten,
``Phases of ${\cal N} = 2$ theories in two dimensions,''
Nucl.\ Phys.\ B {\bf 403}, 159 (1993);
{\tt arXiv:hep-th/9301042}.
}

\lref\ColemanUZ{
S.~R.~Coleman,
``More about the massive Schwinger model,''
Annals Phys.\  {\bf 101}, 239 (1976).
}

\lref\first{
J.~Maldacena and H.~Ooguri,
``Strings in $AdS_3$ and $SL(2,R)$ WZW model. Part 1: the spectrum,''
J.\ Math.\ Phys. 42 (2001) 2929
({\sl Special issue on M Theory}); {\tt arXiv/0001053}.
}

\lref\second{J.~Maldacena, H.~Ooguri and J.~Son,
``Strings in $AdS_3$ and the $SL(2,R)$ WZW model.
Part 2: Euclidean black hole,''
J.\ Math.\ Phys.  42 (2001) 2961
({\sl Special issue on M Theory}); {\tt
arXiv/0005183}.
}

\lref\siw{E.~Silverstein and E.~Witten,
``Criteria for conformal invariance of $(0,2)$ models,''
Nucl.\ Phys.\ B {\bf 444}, 161 (1995);
{\tt arXiv:hep-th/9503212}.
}

\lref\wittenjones{E.~Witten,
``Chern-Simons gauge theory as a string theory,''
{\tt arXiv:hep-th/9207094}.
}

\lref\wittenknot{E. Witten, ``Quantum field theory
and the Jones polynomial,'' Comm. Math. Phys. {\bf 121}, 351
(1989).}

\lref\bcov{M.~Bershadsky, S.~Cecotti, H.~Ooguri and C.~Vafa,
``Kodaira-Spencer theory of gravity and exact
results for quantum string amplitudes,''
Commun.\ Math.\ Phys.\  {\bf 165}, 311 (1994);
{\tt arXiv:hep-th/9309140}.
}


{\Title{\vbox{
\hbox{CALT-68-2428}
\hbox{HUTP-03/A014}
\hbox{\tt hep-th/0302109}}}
{\vbox{
\centerline{The $C$-Deformation of Gluino and Non-planar Diagrams}}}
\vskip .3in
\centerline{Hirosi Ooguri$^1$ and Cumrun Vafa$^{1,2}$}
\vskip .4in
}

\centerline{$^1$ California Institute of Technology 452-48,
Pasadena, CA 91125, USA}
\vskip .1in
\centerline{$^2$ Jefferson Physical Laboratory, Harvard University,
Cambridge, MA 02138, USA}

\vskip .4in
We consider a deformation of ${\cal N}=1$ supersymmetric
gauge theories in four dimensions, which we call the $C$-deformation,
where the gluino field satisfies
a Clifford-like algebra dictated by a self-dual two-form,
instead of the standard Grassmannian algebra. The superpotential
of the deformed gauge theory is computed by the full partition
function of an associated matrix model (or more generally
a bosonic gauge theory), including non-planar diagrams.  In
this identification, the strength of the two-form controls
the genus expansion of the matrix model partition function.
For the case of pure ${\cal N}=1$ Yang-Mills this deformation
leads to the identification of
 the all genus partition function of $c=1$ non-critical bosonic
string at self-dual radius
as the glueball superpotential.
Though the $C$-deformation violates Lorentz invariance,
the deformed $F$-terms are Lorentz invariant and
the Lorentz violation is screened in the IR.

\vfill
\eject

\lref\agnt{
I.~Antoniadis, E.~Gava, K.~S.~Narain and T.~R.~Taylor,
``Superstring threshold corrections to Yukawa couplings,''
Nucl.\ Phys.\ B {\bf 407}, 706 (1993);
{\tt arXiv:hep-th/9212045}.
}

\newsec{Introduction}

Topological strings \wittenjones\ and
its connection to superstrings
\refs{\bcov,\agnt} have proven
to be rather important for a better understanding of
the dynamics of ${\cal N}=1$
supersymmetric gauge theories in four dimensions.  In particular,
the open/closed topological string duality conjectured in
\ref\gova{R.~Gopakumar and C.~Vafa,
``On the gauge theory/geometry correspondence,''
Adv.\ Theor.\ Math.\ Phys.\  {\bf 3}, 1415 (1999),
{\tt arXiv:hep-th/9811131}.
} and proven in
\ref\oova{
H.~Ooguri and C.~Vafa,
``Worldsheet derivation of a large $N$ duality,''
Nucl.\ Phys.\ B {\bf 641}, 3 (2002),
{\tt arXiv:hep-th/0205297}.
} leads to
some non-perturbative predictions for ${\cal N}=1$ gauge theories
in 4 dimensions constructed as low energy
limits of superstring theory \ref\vaug{C.~Vafa,
``Superstrings and topological strings at large $N$,''
J.\ Math.\ Phys.\  {\bf 42}, 2798 (2001),
{\tt arXiv:hep-th/0008142}.
}.  Some of these predictions (coming from genus
0 computations on the closed string side) relate
to the superpotential for the glueball fields
\ref\civ{F.~Cachazo, K.~Intriligator and C.~Vafa,
``A large $N$ duality via a geometric transition,''
Nucl. \ Phys. \ B {\bf 603}, 3 (2001),
{\tt arXiv:hep-th/0103067}.}.  This relation
has recently been better understood and has led
to a striking connection between a wide class of
${\cal N}=1$ supersymmetric gauge theories
and planar diagrams of matrix models (or more generally
the associated bosonic gauge theories) \ref\dv{
R.~Dijkgraaf and C.~Vafa,
``Matrix models, topological strings, and supersymmetric gauge theories,''
Nucl.\ Phys.\ B {\bf 644}, 3 (2002), {\tt
arXiv:hep-th/0206255};
``On geometry and matrix models,''
Nucl.\ Phys.\ B {\bf 644}, 21 (2002),
{\tt arXiv:hep-th/0207106};
``A perturbative window into non-perturbative physics,''
{\tt arXiv:hep-th/0208048};
``N=1 supersymmetry, deconstruction, and bosonic gauge theories,''
{\tt arXiv:hep-th/0302011.}.}.

However
 the open/closed string duality suggests
an even more extensive insight into the dynamics
of ${\cal N}=1$ supersymmetric
gauge theory.  In particular the closed string side
is an ${\cal N}=2$ supersymmetric theory, deformed to
${\cal N}=1$ by turning on fluxes.  The
topological string
computes F-terms for the ${\cal N}=2$ supersymmetric
 closed string dual of the form \refs{\bcov,\agnt}
\eqn\grvi{\int d^4x d^4\theta
({\cal W}_{\alpha \beta}{\cal W}^{\alpha \beta})^{g}
F_g(S_i)}
where ${\cal W}_{\alpha \beta}$ denotes the ${\cal N}=2$ graviphoton
superfield, and the $d^4\theta$ denotes a superintegral over
half of the $8$ super-directions of the ${\cal N}=2$ superspace,
and $S_i$ denote vector multiplets of ${\cal N}=2$.  Let us
write the four ${\cal N}=2$ super-directions of \grvi\ by exhibiting
its ${\cal N}=1$ content
 as $(\theta^{\alpha}, {\hat \theta}^\alpha)$.
As pointed out in \vaug\ turning on fluxes deforms
this to an ${\cal N}=1$ supersymmetric theory by giving
vev to an auxiliary field of $S_i$ of the form
$$S_i(\theta, {\hat \theta})=S_i (\theta )+N_i {\hat \theta^2}$$
where $S_i(\theta)$ can now be viewed as an ${\cal N}=1$
chiral superfield, which in the gauge theory
context will be interpreted as a glueball superfield.
 To write the content of \grvi\ in
purely ${\cal N}=1$ terms we can do one of two things:  We
can absorb two ${\hat \theta}$'s  by expanding ${\cal W}_{\alpha \beta}$
to obtain the ${\cal N}=1$  gravitino multiplet ${\cal W}_{\alpha \beta
\gamma}$, or we can use the auxiliary field vev of the $S_i$ above
to absorb them.  Turning on the graviphoton
field $F_{\alpha \beta}$ (now viewed as a parameter in the ${\cal N}=1$
supersymmetric theory), leads to two terms in the action
$$\eqalign{
&\Gamma_1=g \int d^4x d^2\theta{\cal W}_{\alpha \beta \gamma}
{\cal W}^{\alpha \beta \gamma} (F_{\delta \zeta}F^{\delta \zeta})^{g-1}
 F_g (S_i), \cr
&\Gamma_2=\int d^4x d^2\theta (F_{\alpha \beta}F^{\alpha \beta})^{g}
N_i {\partial F_g\over \partial S_i}.}$$
Note that the first term $\Gamma_1$ exists even
if we do not break the supersymmetry from ${\cal N}=2$
to ${\cal N}=1$.  In particular it is present even
if $N_i=0$.  The second term $\Gamma_2$ is more in
tune with breaking supersymmetry to ${\cal N}=1$.
If we turn off the Lorentz violating term $F_{\alpha \beta}=0$
then we only have the $g=1$ part of $\Gamma_1$, giving
terms of the form $\int d^4x R^2$ (with appropriate index contractions),
or the $g=0$ part of $\Gamma_2$,
giving the superpotential term for the glueball
field.

The main question is to give an interpretation of turning
on the Lorentz violating graviphoton background 
$F_{\alpha \beta}$ in purely
${\cal N}=1$ gauge theoretic terms.\foot{This question
was raised in \vaug\ where it was proposed
that it may be related to making space non-commutative.  Here
we find a different interpretation.}  We will find
a satisfactory answer to this question in this paper.  In particular
we find that deforming the classical anti-commutativity
of the gluino fields by making it satisfy the Clifford
algebra, dictated by the vev of $F_{\alpha \beta}$ of the form,
$$\{ \psi_\alpha , \psi_{\beta} \}=2F_{\alpha \beta} ,$$
leads to the ${\cal N}=1$ realization of the string deformation.
We will see how this arises in string theory and field theory
context.  The string theory derivation follows the
setup of  \bcov\ and the more general field theory derivation
follows the setup introduced in \ref\gri{
R.~Dijkgraaf, M.~T.~Grisaru, C.~S.~Lam, C.~Vafa and D.~Zanon,
``Perturbative computation of glueball superpotentials,''
{\tt arXiv:hep-th/0211017}. 
}.  Even though the field theory argument is
 more general (and in particular includes field theories that are not
known to be constructible in string theory context), the
intuition and ideas coming from the string derivation are crucial
for field theory derivation.
We in particular
find a simple map between the superspace part of these two computations.  This
leads to the statement that the gradient of the full partition function of
the matrix model (not just its planar limit)
with potential equal to the superpotential
of the gauge theory, computes the superpotential of the
associated supersymmetric gauge theory, where the $|F|$ gets
identified with the genus counting parameter of the matrix model.\foot{
This can be generalized to the deformation of the $U(1)$ coupling
constants in a straight-forward manner.}
This completes the interpretation of the meaning of $\Gamma_2$ from
the gauge theory side.  The interpretation of $\Gamma_1$
should follow a similar derivation.

The organization of this paper is as follows:  In section 2 we
motivate and define $C$-deformation by studying string theory
diagrams with graviphoton turned on.  In section 3 we show how
the relevant part of the topological string computation works with
graviphoton turned on.  In section 4 we discuss the field theory
limit and how to obtain the same results using the more general field theory
setup \'a  la \gri .  In section 5 we discuss the physical
interpretation of this deformation.

\newsec{The $C$-deformation}

\lref\ovknot{
H.~Ooguri and C.~Vafa,
``Knot invariants and topological strings,''
Nucl.\ Phys.\ B {\bf 577}, 419 (2000),
{\tt arXiv:hep-th/9912123}.
}

It has been shown in \refs{\bcov,\agnt} that the partition functions
of topological closed string on a Calabi-Yau three-fold $M$
compute the $F$-terms of the four dimensional theory
obtained by compactifying Type II superstring on $M$. A similar
statement holds when we add D branes \refs{\bcov,\ovknot,\vaug}.
The partition
functions of topological string ending on D branes wrapping on
$n$-dimensional cycles on $M$ give the $F$-terms for
the ${\cal N}=1$ supersymmetric gauge theory in four dimensions
which is defined as the low energy limit of Type II superstring
with D($n+3$) branes wrapping on these cycles and extending
in four dimensions. The $F$-terms take the form
\eqn\potential{
    N \sum_{g, h} \int d^4x d^2\theta
   \ ( {\cal W}_{\alpha\beta} {\cal W}^{\alpha\beta})^g \
       h S^{h-1} \ F_{g,h},}
where $N$ is the rank of the $U(N)$ gauge group,
${\cal W}_{\alpha\beta}$ is the supergravity multiplet
whose bottom component is the self-dual part of the graviphoton
field strength
\eqn\sugramultiplet{
{\cal W}_{\alpha\beta} = F_{\alpha\beta} + \cdots ,}
with $\alpha, \beta = 1,2$ being
spinor indices in four dimensions,
$S$ is the glueball superfield,
\eqn\glueballeffectiveaction{
  S = {1\over 32 \pi^2}
\epsilon^{\alpha\beta} {\rm Tr} \ {\cal W}_\alpha {\cal W}_\beta , }
where ${\cal W}_\alpha$ is the chiral superfield with gluino $\psi_\alpha$
as its bottom component,
$$ {\cal W}_\alpha = \psi_\alpha + \cdots .$$
 The topological string computes the
coefficients $F_{g,h}$ as the partition function for
genus $g$ worldsheet with $h$ boundaries.

In particular the terms in \glueballeffectiveaction\ with $g=0$,
namely the sum over the planar worldsheets, gives the effective
superpotential for $S$ as $Nd{\cal F}_0/dS$ where
\eqn\genuszero{ {\cal F}_0(S) =  \sum_h S^h \ F_{0,h}.}
Combining this with the fact \wittenjones\
that the partition function for
the topological string on the D$(n+3)$ brane can be computed
using the Chern-Simons theory (or its dimensional reduction)
 leads to the statement that the effective
action is computed by a sum over planar diagrams of the Chern-Simons
theory, or its reduction to the matrix model
for specific class of D5 branes wrapping 2-cycles \dv.

\lref\dh{
M.~R.~Douglas and C.~M.~Hull,
``D-branes and the noncommutative torus,''
JHEP {\bf 9802}, 008 (1998),
{\tt arXiv:hep-th/9711165}.
}

\lref\sw{
N.~Seiberg and E.~Witten,
``String theory and noncommutative geometry,''
JHEP {\bf 9909}, 032 (1999),
{\tt arXiv:hep-th/9908142}.
}
The purpose of this paper is to understand the
 meaning of the sum over non-planar diagrams. Among terms that are
generated by the flux
we have
\eqn\nonplanaraction{
  \Gamma(S, F) = N \ \sum_g \int d^4 x \
      (F_{\alpha\beta}F^{\alpha\beta})^g \ \int d^2\theta
{\partial \over \partial S} F_g \left(S(\theta)\right) ,}
where
\eqn\genusgamp{ F_g(S)= \sum_h \ S^h \ F_{g,h}.}
This gives the effective action for the glueball superfield $S$
when the graviphoton field strength is non-zero. The question
is whether there is a purely ${\cal N}=1$ gauge theoretical interpretation
of the same quantity without invoking the coupling to the ${\cal N}=2$
supergravity field. Does the graviphoton deform the gauge
theory in a way similar to the Neveu-Schwarz two form $B_{\mu\nu}$
generating noncommutativity of coordinates on D branes
\refs{\dh,\sw}?

\lref\berktend{
N.~Berkovits,
``Super-Poincare covariant quantization of the superstring,''
JHEP {\bf 0004}, 018 (2000),
{\tt arXiv:hep-th/0001035}.
}

\lref\berkrr{
N.~Berkovits,
``Quantization of the superstring in Ramond-Ramond backgrounds,''
Class.\ Quant.\ Grav.\  {\bf 17}, 971 (2000),
{\tt arXiv:hep-th/9910251}.
}

\lref\bv{
N.~Berkovits and C.~Vafa,
``${\cal N}=4$ topological strings,''
Nucl.\ Phys.\ B {\bf 433}, 123 (1995),
{\tt arXiv:hep-th/9407190}.
}

\lref\berkcy{
N.~Berkovits,
``Covariant quantization of the Green-Schwarz superstring
in a Calabi-Yau background,''
Nucl.\ Phys.\ B {\bf 431}, 258 (1994),
{\tt arXiv:hep-th/9404162}.
}

The relation for the topological string amplitudes and
the $F$-terms for the type II string compactification
was originally derived using the NSR formalism, where the
graviphoton vertex operator in the Ramond-Ramond sector
is constructed in terms of the spin field. It was observed
in \bcov\ that it generates the topological twist on the
worldsheet, which gives the connection between the type II
string computation and the topological string computation.
A precise derivation of the connection is rather nontrivial,
involving a sum over spin structure and nontrivial identities
of theta functions \agnt . A more economical derivation was found in \bv\
making use of the covariant quantization of type II superstring
compactified on a Calabi-Yau three-fold, which was developed
in \berkcy .

Compared with the covariant quantization of superstring in
ten dimensions \berktend , the formalism is substantially simpler
for superstring compactified on a Calabi-Yau three-fold since
the supersymmetry we need to make manifest is smaller. In fact,
the four-dimensional part of the worldsheet Lagrangian density
that is relevant for our discussion is simply given by
\eqn\covariant{
 {\cal L} = {1\over 2} \partial X^\mu \bar \partial X_\mu
  + p_\alpha \bar \partial \theta^\alpha +
p_{\dot \alpha} \bar \partial \theta^{\dot \alpha}+
 \bar{p}_\alpha
\partial \bar \theta^\alpha + \bar p_{\dot \alpha}
\partial \bar \theta^{\dot \alpha},}
where $p$'s are $(1,0)$-forms, $\bar p$'s are
$(0,1)$-forms, and $\theta ,\bar \theta$'s are $0$-forms.
The remainder of the Lagrangian density consists of the topologically
twisted ${\cal N}=2$
supersymmetric sigma-model on the Calabi-Yau three-fold and
a chiral boson which is needed to construct the R current.
It is useful to note that the worldsheet theory defined by
\covariant\ (excluding the fermionic fields
with dotted indices)
can be regarded as a topological B-model on $R^4$.
The topological BRST symmetry is defined by
\eqn\topBRST{ \eqalign{ & \delta X_{\alpha\dot\alpha}
= \epsilon_{\dot \alpha} \theta_\alpha +
\bar \epsilon_{\dot \alpha} \bar\theta_{\alpha} ,\cr
& \delta \theta_\alpha = 0,~~ \delta \bar \theta_{\alpha} = 0, \cr
& \delta p_\alpha =  \epsilon^{\dot \alpha}\partial X_{\alpha \dot \alpha}
,~~ \delta \bar p_\alpha =
 \bar \epsilon^{\dot \alpha}\bar \partial X_{\alpha \dot \alpha} ,}}
where $X_{\alpha\dot \alpha} = \sigma_{\alpha \dot \alpha}^\mu X_\mu$
with $\sigma^0=-1$ and $\sigma^{1,2,3}$ being the Pauli matrices, and
raising and lowering of spinor indices are done by using
the anti-symmetric tensors $\epsilon_{\alpha\beta},
\epsilon_{\dot \alpha \dot \beta}$ as usual. We recognize
that this topological symmetry is closely related to the
spacetime supersymmetry. In fact, modulo some shift of $X_{\alpha
\dot \alpha}$ by $\theta_\alpha \theta_{\dot \alpha}$
and $\bar \theta_\alpha \bar \theta_{\dot \alpha}$, \topBRST\ is
identical to transformations generated by the anti-chiral
components $Q_{\dot \alpha}, \bar Q_{\dot \alpha}$
of the ${\cal N}=2$ supercharges in the bulk.

When the worldsheet is ending on D branes and extending in four
dimensions, the boundary conditions for the worldsheet variables
are given by
\eqn\boundarycondition{\eqalign{
  & (\partial - \bar\partial) X^\mu = 0, \cr
  & \theta^\alpha = \bar \theta^\alpha, ~~p_\alpha = \bar p_\alpha.}}
Here we assume that the boundary is located at ${\rm Im}\ z=0$.
These boundary conditions preserve one half of the supersymmetry
generated by $Q+ \bar Q$.

Let us turn on the graviphoton field strength $F_{\alpha\beta}$
and the gluino superfield ${\cal W}_\alpha$, both of which we assume
to be constant. They couple to the bulk and the boundaries
of the string worldsheet as
\eqn\deformation{
\int  \ F^{\alpha\beta} {\cal J}_\alpha \bar {\cal J}_{\beta}
+ \oint {\cal W}^\alpha {\cal J}_\alpha, }
where  ${\cal J}_\alpha, \bar
{\cal J}_\beta$
are the worldsheet currents for the spacetime supercharges $Q_\alpha,
\bar Q_\beta$ \berkrr . We find it convenient to work in the chiral
representation of supersymmetry,\foot{Our convention in this
paper is related to that of \refs{\bv,\berkcy} by redefinition of the
worldsheet variables by 
$$\eqalign{ p_\alpha& \rightarrow p'_\alpha
 = p_\alpha -i\theta^{\dot \alpha}
\partial X_{\alpha \dot \alpha} - {1\over 4} \theta^2
\partial \theta_\alpha, \cr
  p_{\dot \alpha} &\rightarrow
p_{\dot \alpha}' = p_{\dot \alpha} + i \theta^\alpha
\partial X_{\alpha\dot\alpha} -{1\over 2} \theta^2 \partial \theta_{\dot\alpha}
+ {1\over 4} \theta_{\dot \alpha} \partial \theta^2, \cr
 X_{\alpha\dot\alpha}& \rightarrow X_{\alpha\dot\alpha}'
= X_{\alpha\dot\alpha} + i \theta_\alpha \theta_{\dot\alpha}
+i\bar\theta_\alpha \bar\theta_{\dot\alpha}.}$$
\lref\CornalbaCU{
L.~Cornalba, M.~S.~Costa and R.~Schiappa,
``D-brane dynamics in constant Ramond-Ramond 
potentials and  noncommutative geometry,''
{\tt arXiv:hep-th/0209164}.
}
See also \CornalbaCU\ for a related discussion. }
in which they are given by
\eqn\chiralsuper{\eqalign{
& {\cal J}_\alpha =   p_\alpha , \cr
& {\cal J}_{\dot \alpha} =
 p_{\dot \alpha} -2i \theta^\alpha \partial
X_{\alpha \dot \alpha} + \cdots ,}}
where $\cdots$ in the second line represents terms
containing $\theta^{\dot \alpha}$ and
$\theta^2 = \epsilon_{\alpha\beta}\theta^\alpha \theta^\beta$.
The second set of supercharges $\bar Q_\alpha, \bar Q_{\dot \alpha}$
are defined by replacing $p, \theta$ by
$\bar p, \bar\theta$.
In this convention, the coupling \deformation\ of the graviphoton
and the gluino to the worldsheet becomes
\eqn\coupling{ \int F^{\alpha\beta}p_\alpha \bar p_\beta
+ \oint {\cal W}^\alpha p_\alpha .}
This simplifies our analysis in this section.
In the field theory limit, the supercharges in this convention
take the form,
\eqn\supercharge{
\eqalign{ & Q_\alpha = {\partial \over \partial \theta^\alpha}
,\cr
& Q_{\dot \alpha} = {\partial \over \partial \theta^{\dot \alpha}}
+ 2i \theta^{\alpha} {\partial \over
\partial x^{\alpha \dot \alpha}}
.}}
The boundary conditions \boundarycondition\
identify the supercurrents ${\cal J} = \bar {\cal J}$,
reducing the supersymmetry to ${\cal N}=1$.

\subsec{Deformed superspace}

Let us analyze the effect of the graviphoton in the bulk.
We will find it useful to keep track of mass dimensions
of operators, so we introduce the string scale $\alpha'$,
which has dimension $-2$. As usual $\theta^\alpha$ has
dimension $-1/2$ and its conjugate $p_\alpha$ has
dimension $+1/2$. The gluino
${\cal W}_\alpha$ has dimension $+3/2$. For the graviphoton
field strength $F_{\alpha\beta}$, we assign dimension $+3$.
One might have thought that dimension $+2$ would be canonical
for the field strength. Here we assign dimension $+3$
to $F_{\alpha\beta}$ so that the higher genus contributions
to the superpotential \nonplanaraction\ remain finite
in the field theory limit $\alpha' \rightarrow 0$. For example,
as we will see later, the genus $g$ contribution to the superpotential
in the pure ${\cal N}=1$ super Yang-Mills theory is of
the form $\sim N \ (F_{\alpha\beta}F^{\alpha\beta})^g \ S^{2-2g}$ and it has
 dimension $3$ (which is the correct dimension for
the superpotential in four dimensions) for all $g$ only if we
assign the same dimension to $F$ and $S = {1\over 32\pi^2}
\epsilon^{\alpha\beta}
{\rm Tr} \  {\cal W}_\alpha {\cal W}_\beta$.
With this assignment of mass dimensions, the relevant
part of the Lagrangian density is expressed as
\eqn\withFnoW{
 {\cal L}={1 \over 2\alpha'} \partial X^\mu \bar\partial X_\mu
 + p_\alpha \bar \partial \theta^\alpha + \bar p_\alpha \partial
  \bar \theta^\alpha + \alpha'^2 F^{\alpha \beta} p_\alpha
  \bar p_\beta.}
As we mentioned, we are working in the chiral representation
where supercharges are defined by \chiralsuper .

It is useful to note that the self-dual configuration of the
graviphoton, namely
\eqn\selfdual{
F_{\alpha\beta} \neq 0, ~~ F_{\dot \alpha \dot \beta} = 0,}
gives an exact solution to the string equation of motion.
This can be seen clearly from the fact that the perturbed
action \withFnoW\ does not break the
conformal invariance on the worldsheet. From the target space
point of view, we see that the energy-momentum tensor for
the graviphoton vanishes for \selfdual , so there is no
back-reaction to the metric.\foot{The energy-momentum
tensor for the non-zero field strength can vanish since
the self-dual field strength becomes complex valued when
analytically continued to Minkowski space.}

\lref\vann{J.H. Schwarz and P. van Nieuwenhuizen, ``Speculations
concerning a fermionic structure of space-time,'' Lett. Nuovo Cim.
{\bf 34},21 (1982).}
\lref\ferr{S. Ferrara and M.A. Lledo, ``Some aspects of deformations
of supersymmetric field theories,'' JHEP {\bf 0005}, 008 (2000),
{\tt arXiv:hep-th/0002084}.}
%
\lref\KlemmYU{
D.~Klemm, S.~Penati and L.~Tamassia,
``Non(anti)commutative superspace,''
{\tt arXiv: hep-th/0104190}.
}

Now let us add boundaries to the worldsheet.
For the moment, we turn off the gluino field $\psi_\alpha=0$
and discuss effects due to the graviphoton field strength.
In the presence of graviphoton, the equations of motion for
$\theta$ and $\bar \theta$ are deformed to
\eqn\deformedeq{
\eqalign{ & \bar \partial \theta^\alpha = \alpha'^2 F^{\alpha\beta}
             \bar p_\beta \cr
   & \partial \bar \theta^\alpha = - \alpha'^2 F^{\alpha\beta} p_\beta.}}
Before turning on $F^{\alpha\beta}$, the only nontrivial
operator product is that between $p_\alpha$ and $\theta^\alpha$
(and between $\bar p_\alpha$ and $\bar\theta^\alpha$) as
\eqn\pthetaope{ p_\alpha(z) \theta^\beta(w) \sim {\delta_\alpha^\beta \over
   2\pi i (z-w)}.}
The relation \deformedeq\ modifies this. If we write
$$ \theta^\alpha = \Theta^\alpha + \eta^\alpha, ~~
   \bar \theta^\alpha = \Theta^\alpha - \eta^\alpha,$$
so that the boundary condition is $\eta=0$, the relation
\deformedeq\ together with the short distance
singularity \pthetaope\ imply
\eqn\ope{
\eqalign{ \Theta^\alpha(z) \Theta^\beta(w) & \sim
 {1\over 2\pi i} \alpha'^2 F^{\alpha\beta}
\log\left[ {z-\bar w \over \bar z - w} \right] , \cr
\Theta^\alpha(z) \eta^\beta(w) & \sim -
{1 \over 4\pi i} \alpha'^2 F^{\alpha\beta}
\log \left[
{(z-w)(\bar z - \bar w) \over
(z-\bar w) (\bar z - w) } \right], \cr
\eta^\alpha(z) \eta^\beta(w) & \sim 0.}}
In particular, on the boundary we have
$$ \theta^\alpha(\tau+\epsilon)
 \theta^\beta(\tau)
 + \theta^\beta(\tau) \theta^\alpha(\tau-\epsilon) =
 2\alpha'^2 F^{\alpha\beta}. $$
Therefore a correlation function of
operators $\theta=\bar\theta$ on the boundary
with the time-ordering along the boundary
obey the Clifford algebra
\eqn\clifford{
 \{ \theta^\alpha , \theta^\beta \} = 2 \alpha'^2 F^{\alpha\beta}, }
rather than the standard Grassmannian algebra.
Such deformation of the superspace has been studied
earlier \refs{\vann,\ferr,\KlemmYU}, and it is interesting
that it is realized in the context of string theory.\foot{
The result of this subsection
has been generalized to other dimensions in a recent work
\ref\neww{J.~de Boer, P.~A.~ Grassi, P.~ van Nieuwenhuizen,
``Non-commutative superspace from string theory,''
{\tt arXiv:hep-th/0302078}.}.}

The presence of the factor $\alpha'^2$ in \clifford\
means that the deformation of the superspace
 does not survive the field theory limit
$\alpha' \rightarrow 0$ unless we simultaneously
take $F^{\alpha \beta} \rightarrow \infty$ so that
$\alpha'^2 F^{\alpha\beta}$ remains finite. It may
be possible to make sense of such a limit since
the constant graviphoton field strength is an exact
solution to the string equation of motion for any
large value of $F_{\alpha\beta}$ as we explained
earlier. It turns out, however, if one wants to preserve
the ${\cal N}=1$ supersymmetry, we will need
to make another modification to the theory, which
we call the $C$-deformation. We will find that this
restores the anticommutativity of $\theta$'s and
undoes the deformation of the superspace. These 
effects survive the field
theory limit without taking $F^{\alpha\beta}$ to be large.

\subsec{$C$-deformation of gluino and undeforming of superspace}

Since we work in the chiral representation where $Q_\alpha
=\oint p_\alpha $, the supercharges in the field theory
limit takes the form,
$$ \eqalign{ &Q_\alpha = {\partial \over \partial \theta^\alpha} \cr
& Q_{\dot \alpha} = {\partial \over \partial \theta^{\dot \alpha}}
+2i \theta^\alpha {\partial \over \partial x^{\alpha\dot \alpha}}.}$$
The deformation of the superspace by \clifford\ would then
modify the supersymmetry algebra as
\eqn\deformedsusy{ \eqalign{
& \{ Q_\alpha, Q_{\dot \beta} \} = 2i {\partial \over \partial
x^{\alpha\dot \beta}}, \cr
& \{ Q_\alpha, Q_\beta\} = 0, \cr
& \{ Q_{\dot \alpha}, Q_{\dot \beta} \}
= -8 \alpha'^2 F^{\alpha\beta} {\partial^2
\over \partial x^{\alpha\dot\alpha} \partial x^{\beta\dot\beta}}.}}

A closely related issue arises on the string worldsheet,
where the constant graviphoton field strength breaks supersymmetry
on the D branes. When $F_{\alpha\beta}=0$,
there are two sets of supercharges $Q$ and $\bar Q$, which
are identified on the boundary $Q=\bar Q$ by the boundary conditions
\boundarycondition .
It turns out that the graviphoton vertex operator $F^{\alpha\beta}
p_\alpha \bar p_\beta$ is not invariant under the supersymmetry
but transforms into a total derivative on the worldsheet.
Let us consider the combination $\epsilon^{\dot \alpha}
\left( Q_{\dot \alpha} + \bar Q_{\dot \alpha}\right)$, which
preserves the boundary conditions. We find
\eqn\boundarybreaking{
\eqalign{  \delta \left[
 \alpha'^2 \int  \ F^{\alpha\beta} p_\alpha \bar p_\beta \right]
& = 2i \alpha' \epsilon^{\dot \alpha} F^{\alpha\beta} \int
d \left[   Y_{\alpha \dot \alpha}
    (p_\beta+ \bar p_\beta) \right] \cr
&= \sum_{i=1}^h 4\alpha' \epsilon^{\dot \alpha}
F^{\alpha\beta} \oint_{\gamma_i} Y_{\alpha \dot \alpha}\
 p_\beta ,}}
where
\eqn\whaty{ Y_{\alpha\dot \beta} = X_{\alpha\dot\beta} +
i \theta_\alpha \theta_{\dot \alpha} + i \bar\theta_\alpha
\bar \theta_{\dot \alpha}.}
and $\gamma_i$'s are boundaries
of the worldsheet. Therefore, as it is, the
supersymmetry is broken on the boundaries of the
worldsheet. Unlike the deformation of the superspace
\clifford , which disappears in the field theory limit
$\alpha'\rightarrow 0$, the amount of supersymmetry
breaking is comparable to the gluino coupling
$\alpha' \oint {\cal W}^\alpha p_\alpha$ and
therefore is not negligible in this limit.

On the other hand, if the gluino fields ${\cal W}_\alpha$
are constant Grassmannian variables taking value in the diagonal
of the $U(N)$ gauge group, its coupling to the worldsheet
does not break the topological invariance since
\eqn\gluinovariation{
\delta \oint_{\gamma} {\cal W}^\alpha p_\alpha
=2i\epsilon^{\dot \alpha}{\cal W}^\alpha
\oint_{\gamma} d Y_{\alpha
\dot \alpha} =0. }
It turns out that
there is a natural way to modify this assumption so that
the variation of the gluino coupling precisely
cancels the boundary term generated by the graviphoton
in the bulk. That is to assume that the gluino fields make
the Clifford algebra
\eqn\cdeform{
 \{ \psi_\alpha, ~\psi_\beta \} = 2F_{\alpha \beta}. }
Note that the mass dimensions of the both sides of this
equation match up without introducing the string scale
$\alpha'$, so this relation survives the field theory limit
$\alpha' \rightarrow 0$ without making $F$ large. In the
following computation, we continue to assume that
$\psi_\alpha$ is constant and takes value in the
diagonal of $U(N)$. In a more general situation, we
interpret \cdeform\ as saying that
\eqn\cdeformtwo{
\{ {\cal W}_{\alpha}(x), ~{\cal W}_{\beta}(x)\}^i_{~j}
 = F_{\alpha\beta}\ \delta^i_{~j}~~~{\rm mod}~ \bar D,} 
where $i,j=1,\cdots,N$ and the product
${\cal W}_\alpha {\cal W}_\beta$ in the left-hand side
includes the matrix multiplication
with respect to these $U(N)$ indices. Note that the
identity is modulo 
$D_{\dot \alpha}$ 
since that is all we need to cancel the boundary term.
Therefore \cdeformtwo\ should be regarded as a relation 
in the chiral ring. We call this the $C$-deformation of 
the gluino.

This deformation also undoes the deformation of the superspace \clifford .
The analysis of the previous section changes because the
gluino is turned on, and it affects the boundary condition 
of $\theta$ and $p$. One can easily show that the $C$-deformation
of the gluino compensates the effect of the 
graviphoton on correlation functions of $\theta$'s on
the boundaries and restores the anticommutativity
of $\theta$'s there. Namely, $\theta$'s remain ordinary
Grassmannian variables and the superspace is undeformed. 
This eliminates the $F^{\alpha\beta}$ dependent term in 
\deformedsusy\ and recovers the standard supersymmetry algebra. 
This is consistent with the fact that the $C$-deformation
of the gluino restores the spacetime supersymmetry in the
graviphoton background. 

Let us show that the $C$-deformation cancels the boundary terms
\boundarybreaking\ and restores the supersymmetry.
Since $\psi$'s do not anti-commute with each other, we
need to use the path-ordered exponential,
\eqn\boundaryfactor{{\cal P}\exp \left( \alpha' \oint_\gamma
{\cal W}^\alpha p_\alpha  \right),}
along each boundary to define the gluino coupling.  As we will
see below such a term makes sense, $i.e.$ it
does not depend on the origin of the path-ordering,
 as long as the $\oint p_\alpha$
through each boundary is zero.  Let us
 evaluate the variation of the boundary factor
\boundaryfactor\ and find
\eqn\almost{
\eqalign{ & \delta\left[
{\cal P} \exp \left( \alpha' \oint_\gamma {\cal W}^\alpha
p_\alpha  \right) \right] \cr
&= 2i\epsilon^{\dot \alpha}{\cal P}
 \left[  \oint_\gamma {\cal W}^\alpha d Y_{\alpha\dot \alpha}
 \exp \left( \alpha'\oint_\gamma {\cal W}^\alpha
p_\alpha   \right) \right] \cr
& = -2i \alpha'  \epsilon^{\dot \alpha}F^{\alpha\beta}
{\cal P} \left[ \left(\oint_\gamma Y_{\alpha \dot \alpha}
p_\beta - Y_{\alpha\dot \alpha}(o)
\oint_\gamma p_\beta \right)
 \exp \left( \alpha'\oint_\gamma {\cal W}^\alpha
p_\gamma \right)\right].}}
Here $o$ is an arbitrarily chosen base point on the boundary
$\gamma$ which is used to define the path-ordering.

This almost cancels the boundary terms \boundarybreaking\
coming from the graviphoton variation, except for the term
$Y_{\alpha\dot\alpha}(o) \oint p_\beta $, which depends on the choice
of the base point $o$.  If $\oint p_\beta$ through
each boundary is zero, this definition of path-ordering is
independent of the base point $o$, and its supersymmetry
variation completely cancels the graviphoton variation.
For the worldsheet
with a single boundary, the condition that $\oint p_\beta$ vanish
is automatic, as the boundary is homologically trivial.
If there are more boundaries $h>1$,
we need to insert an operator which
enforces $\oint p_\beta =0$ to make the path-ordering
well-defined. In particular,
the dependence on the base point of path-ordering
disappears if we compute a correlation function of $2(h-1)$
gluino fields, which inserts
$$\prod_{i=1}^{h-1} \left(\alpha'
\oint_{\gamma_i} {\cal W}^\alpha p_{\alpha}
\right)^2 = \left(\alpha'^2
\epsilon^{\alpha\beta}{\cal W}_\alpha{\cal W}_\beta
\right)^{h-1} \ \prod_{i=1}^{h-1} \epsilon^{\alpha\beta}
\oint_{\gamma_i} p_\alpha \oint_{\gamma_i} p_\beta.$$
Note that for these insertion, which are {\it not} path-ordered, the
Grassmannian property of $p_{\alpha}$ projects the ${\cal W}$ contribution
on the antisymmetric part via $\epsilon_{\alpha \beta}$.
There are no contributions from the graviphoton $F^{\alpha\beta}$,
because $\epsilon^{\alpha\beta}F_{\alpha\beta}=0$. Note that
$2(h-1)$
is the maximum number of gluino insertions we can make for a
given number of boundaries if we take into account
the global constraint
$$ \sum_{i=1}^h \oint_{\gamma_i} p_\alpha (\tau)
= {1\over 2} \int d( p_\alpha + \bar p_\alpha)
= 0 . $$
Since $p_\alpha, \bar p_\alpha$ are fermionic, inserting
$2(h-1)$ gluino fields amounts to imposing a constraint
\eqn\momentumconstraint{
 \oint_{\gamma_i} p_\alpha(\tau) = 0,}
on each boundary and the $o$ dependent terms in \almost\ vanishes
in this case, completely cancelling \boundarybreaking .
Therefore we can compute the topological open string amplitude
for worldsheet with $h$ boundaries
if and only if we compute the correlation function of
$2(h-1)$ gluino superfields,
consistently with the structure in \potential .
In other words, the only $F$-terms that make sense in this
context involve insertions of $(h-1)$ factors of $S$.  As we
will see in the next section, the path ordering of the gluino
vertex on all the boundaries leads in the path-integral
computation to a term involving $(F^2)^g$.

The fact that we can make sense of only such $F$-term amplitudes,
which impose the vanishing of the fermionic momentum through each
hole, strongly suggests that the rest of the amplitudes
should be set to $0$.  In some sense, those would be the analog
of trying to obtain a non-gauge invariant correlator in a gauge
invariant theory and finding it to be zero after integrating
over the gauge orbit.

We have found that the string theory computation
in the presence of the constant graviphoton field
strength preserves the topological invariance on
the worldsheet and compute the $F$-terms
\potential\ of the low energy effective theory on
the D branes if we turn on the $C$-deformation
\clifford\ of the gluino fields. The $C$-deformation
also restores the anticommutativity of $\theta$'s,
and thereby undeforms the superspace.

\newsec{Topological string amplitudes}

We found that the constant graviphoton background
by itself does not preserve the ${\cal N}=1$ supersymmetry
on the D branes. We need to turn on the $C$-deformation 
of the gluino \cdeform\ in order to restore the supersymmetry. We can then
compute the $F$-terms for this background
by evaluating the topological string amplitude.
The mechanism to absorb the zero modes of the
worldsheet fermions works essentially in the same way
as described in \refs{\bcov,\bv} in the case of the closed string.
The  only nontrivial part of the topological string computation
is the one that involves the zero modes of
$(p_\alpha, \theta^\alpha)$ system.

Before evaluating the zero mode integral, it
is useful to establish the following formula
on a genus-$g$ worldsheet with $h$ boundaries,
\eqn\zeromodeintegral{
\eqalign{ &
\prod_{i=1}^{h-1} \alpha'^2
\epsilon^{\alpha\beta} \oint_{\gamma_i} p_\alpha
\oint_{\gamma_i} p_\beta \ \times
\exp\left[ \alpha'^2 F^{\alpha\beta}\int  p_\alpha \bar p_\beta
+ \alpha' \sum_{i=1}^h \oint_{\gamma_i} {\cal W}^\alpha
p_\alpha  \right] \cr
&= \prod_{i=1}^{h-1} \alpha'^2
\epsilon^{\alpha\beta} \oint_{\gamma_i} p_\alpha
\oint_{\gamma_i} p_\beta \ \times
 \exp\left[ \alpha'^2 F^{\alpha\beta} \sum_{a,b=1}^{2g}
c^{ab}  \oint_a (p_\alpha + \bar p_\alpha)
\oint_b (p_\alpha  + \bar p_\alpha )\right] .}}
Here we assume that $p_\alpha$ is holomorphic and $\bar p_\alpha$
is anti-holomorphic.
The factor $\prod_i \epsilon^{\alpha\beta} \oint_{\gamma_i} p_\alpha
\oint_{\gamma_i} p_\beta$ is due to the insertion of the
glueball superfield $S^{h-1}$. The gluino fields ${\cal W}_\alpha$
in the left-hand side are
$C$ deformed as in \cdeform ,
and $\exp\left(\oint {\cal W}^\alpha p_\alpha\right)$ is
defined by the path-ordering.
 We choose homology cycles
on the surface as $\gamma_i$ homologous to the
boundaries and
$a,b=1,\cdots,2g$ associated to the handles on the worldsheet,
and $c^{ab}$ is the intersection matrix of these cycles.
Since $c^{ij}=0, c^{ai}=0$, we can add $\gamma_i$ to $a,b$
without changing the intersection number.
This does not change the value of the exponent thanks to
the constraint $\oint_{\gamma_i} p_\alpha =0$
imposed by the insertion of $
\epsilon^{\alpha\beta} \oint_{\gamma_i} p_\alpha \oint_{\gamma_i}
p_\beta$.

The proof of the identity \zeromodeintegral\ consists of two parts.
Let us first evaluate $\exp\left[ \alpha' \oint_\gamma
\psi^\alpha       p_\alpha
\right]$ with the path-ordering along a boundary $\gamma$.
If $\psi_\alpha$ were ordinary Grassmannian variables,
this would be equal to $1$ due to the momentum constraint
$\oint_\gamma p_\alpha = 0$ on the boundary. To obtain a
nontrivial answer, we need to use the anti-commutation
relation \cdeform . For example,
$${\cal P} \oint {\cal W}^\alpha p_\alpha
\oint {\cal W}^\beta p_\beta =2F^{\alpha\beta} \oint p_\alpha(\tau)
\int_o^{\tau} p_\beta ,$$
where we used $\oint p_\alpha =0$ in the last line.
By iteratively using this identity, we can show
\eqn\ordering{
{\cal P} \exp\left( \alpha' \oint_\gamma {\cal W}^\alpha p_\alpha \
\right) = \exp\left[ 2 \alpha'^2 F_{\alpha\beta}
\oint_\gamma p_\alpha(\tau)
\int_o^{\tau} p_\beta\right].}

The second part of the proof is essentially the
same as the proof of the Riemann bilinear identity. We start by writing
\eqn\bilinear{F^{\alpha\beta}
 \int  p_\alpha \bar p_\beta =-{1\over 2}F^{\alpha\beta}
\int d \left[ (p_\alpha(z) + \bar p_\alpha(\bar z))
 \int^z (p_\beta + \bar p_\beta) \right].}
We then cut the worldsheet open along the cycles $a,b$ and also
introduce cuts $\widehat{\gamma}_{i, i+1}$ between the
boundaries $\gamma_i$
and $\gamma_{i+1}$  so that we can perform
the integration-by-parts in the resulting contractible domain (see Fig. 1).
The surface integral in \bilinear\ is then transformed into
contour integrals along the homology cycles $a,b$, the
boundaries $\gamma_i$ and the cuts $\widehat\gamma_{i,i+1}$ connecting them.
The integral along $\widehat{\gamma}_{i,i+1}$ vanishes since
it intersects only with the boundaries $\gamma_i$ and $\gamma_{i+1}$ and
$\oint p_\alpha$ vanishes for them. Thus we are left with
\eqn\bilinearmore{\eqalign{
F^{\alpha\beta}
 \int  p_\alpha \bar p_\beta
&=-{1\over 2} F^{\alpha\beta}\sum_{a,b=1}^{2g}
c^{ab} \oint_a (p_\alpha + \bar p_\alpha )
\oint_b (p_\beta  + \bar p_\beta) \cr
&~~~~~~~-2F^{\alpha\beta}\sum_{i=1}^{h-1} \oint_{\gamma_i}
 p_\alpha(\tau) \int^{\tau}_{o_i} p_\beta
. }}
Combining \bilinear\ with \ordering , we obtain
\zeromodeintegral .

\bigskip\bigskip
\centerline{\epsfxsize 3truein \epsfysize 1.8truein\epsfbox{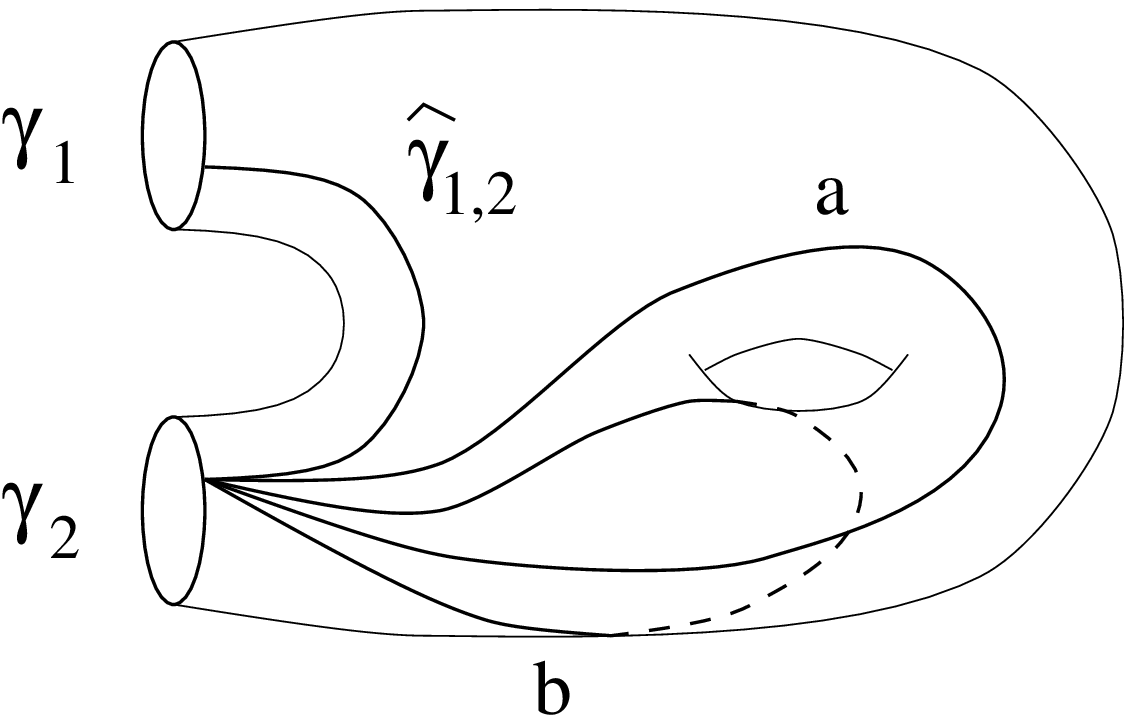}}
\noindent{\ninepoint\sl \baselineskip=8pt {\bf Figure 1}: {\sl
The worldsheet can be made into a contractible
region by cutting along cycles $a,b$ and ${\hat \gamma}_{1,2}$. }}
\bigskip

Let us now evaluate the right-hand side of \zeromodeintegral\
with an explicit parametrization of the fermion zero modes.
A genus-$g$ worldsheet $\Sigma$  with $h$ handles can be constructed
from a genus $(2g+h-1)$ surface $\widetilde\Sigma$ with a complex conjugation
involution $Z_2$ as $\Sigma=\widetilde\Sigma/Z_2$, where the
boundaries of $\Sigma$ are made of $Z_2$ fixed points of $\widetilde
\Sigma$. With respect to the $Z_2$ involution, we can choose
the canonical basis of one forms on $\widetilde{\Sigma}$
as $\{ \omega_a, \tilde\omega_i\}_{a=1,\cdots,2g;~ i=1,\cdots, h-1}$
so that
$$\omega_a(z) = \bar \omega_{a+g}(\bar w)_{|\bar w=z},~~
\tilde\omega_i(z) = \bar{ \tilde\omega}_i(\bar w)_{|\bar w=z}, $$
and normalized as $\oint_{\gamma_i}\omega_j = \delta_{ij}$.
We can then parametrize $p_\alpha$ as
\eqn\pparametrization{
\eqalign{&
p_\alpha = \sum_{a=1}^g \left( \pi^a \omega_a
+ \bar \pi^a \omega_{a+g}\right)
  + \sum_i \tilde\pi^i \tilde\omega_i, \cr
& \bar p_\alpha =
\sum_{a=1}^g \left( \bar \pi^a \bar \omega_a
+ \pi^a \bar \omega_{a+g}\right)
+ \sum_i \tilde\pi^i \bar{\tilde\omega}_i .}}
The right-hand side of \zeromodeintegral\ is then expressed as
\eqn\parametrization{
\eqalign{
&
\prod_{i=1}^{h-1} \alpha'^2
\epsilon^{\alpha\beta} \oint_{\gamma_i} p_\alpha
\oint_{\gamma_i} p_\beta  \times \
\exp\left( \alpha'^2 F^{\alpha\beta}\int  p_\alpha \bar p_\beta
+ \alpha' \sum_{i=1}^h \oint_{\gamma_i} {\cal W}^\alpha
p_\alpha  \right) \cr
&= \prod_{i=1}^{h-1}\alpha'^2
 \epsilon^{\alpha\beta} \tilde\pi_{i\alpha}
\tilde\pi_{j\beta}\ \times \cr
&~~~\times \exp\left[ 2\alpha'^2 F^{\alpha\beta}
 (\pi^a_{\alpha} + \bar \pi^a_{\alpha})\Big(
\left(\Omega_{ab} - \bar \Omega_{a+g,b}\right)\pi^b_{\beta}
+ \left(\bar \Omega_{ab} - \Omega_{a+g,b}\right)
\bar\pi^b_{\beta} \Big) \right] .}}
We can then integrate over the zero modes $\pi_A, \tilde \pi_i$
and obtain
\eqn\final{\eqalign{
&\left\langle
\prod_{i=1}^{h-1} \epsilon^{\alpha\beta} \oint_{\gamma_i} p_\alpha
\oint_{\gamma_i} p_\beta \ \times
\exp\left[ \alpha'^2 F^{\alpha\beta}\int  p_\alpha \bar p_\beta
+ \alpha' \sum_{i=1}^h \oint_{\gamma_i} {\cal W}^\alpha
p_\alpha  \right] \right\rangle \cr
&= \alpha'^{2(2g+h-1)}\ (F_{\alpha\beta} F^{\alpha\beta})^g \
\Big[{\rm det} \ {\rm Im}\left( \Omega_{ab} + \Omega_{a+g,b}\right)
\Big]^2
.}}
The factor $\alpha'^{2(2g+h-1)}$ and the determinant of the period matrix
are cancelled by the integral over the bosonic zero modes
of $X_{\alpha\dot\alpha}$, and we are simply left
with $(F_{\alpha\beta}F^{\alpha\beta})^g$. We have found that the
contribution from the four-dimensional part of the worldsheet
theory is to supply the genus counting factor
$(F_{\alpha\beta}F^{\alpha\beta})^g$ in addition
to the standard $S^{h-1}$ term. All the nontrivial
$g$ and $h$ dependence of $F_{g,h}$ in the $F$-term
should come from the topological string computation for
the internal Calabi-Yau space described by a $\hat{c}=3$, ${\cal N}=2$
superconformal field theory. This is consistent
with the general statement \bcov\ about the
correspondence between the topological string
amplitudes for the $\hat{c}=3$ superconformal field theory
and the $F$-term computation for the Calab-Yau compactification.
Here we have shown explicitly that it works perfectly in the case
of the open string theory if we take into account the $C$-deformation
of the gluino that is necessary to preserve the supersymmetry
in the graviphoton background.

\newsec{Field theory limit}

The field theoretic computation of ${\cal N}=1$
glueball superpotential was performed in \gri\ using
a suitable chiral superspace diagram technique developed in
\ref\grza{M.~T.~Grisaru and D.~Zanon,
``Covariant supergraphs. 1. Yang-Mills theory,''
Nucl.\ Phys.\ B {\bf 252}, 578 (1985).
}.  Let us briefly recall the relevant
part of the computation:  As in \gri ,
we consider the computation in the context
of an adjoint $U(N)$ matter, with some superpotential,
though the generalization to arbitrary cases admitting
large $N$ description is straight-forward.
  One takes an anti-chiral superpotential
${\overline m} {\bar \Phi}^2$ and integrates the ${\bar \Phi}$
out to obtain a theory purely in terms of $\Phi$, given by
$$S=\int d^4xd^2\theta \ \Phi
{1\over 2 \bar m} \left( {\nabla}^2 +{\cal W}^\alpha D_\alpha
\right)
 \Phi +
 W(\Phi)$$
where $W(\Phi)$ is the superpotential, $\nabla^2$ is the ordinary
Laplacian.  In the derivation of this result, it was
assumed that ${\cal W}_{\alpha}$ is covariantly constant, $i.e.$
constant in spacetime and in an Abelian configuration taking
value in the Cartan subalgebra.
Moreover $D_{\alpha} {\cal W}^\alpha=0$.  By integrating
the $\Phi$ out, one obtains an effective superpotential
for the glueball field $S={1\over 32\pi^2} \epsilon_{\alpha \beta}
{\rm Tr} \ {\cal W}^{\alpha}{\cal W}^{\beta }$.  The Feynman
diagrams are dictated by the interaction of $W(\Phi)$, from which
one extracts the ${1\over 2} m\Phi^2$ term and puts it in
 the propagator as usual.
For each internal line $I$ of the Feynman diagram, we have
a propagator given by
\eqn\propa{\int_0^\infty ds_I \ {\rm exp}\left[-{s_I\over 2\bar m}(P_I^2
+{\cal W}^{\alpha}\pi^I_\alpha +
m{\bar m})\right] ,}
where $s_I$ denotes the Schwinger time, $P_I$ and $\pi^I$ are the bosonic
and fermionic
momenta along the line.
Moreover ${\cal W}^{\alpha}$ acts as an adjoint action on the
boundaries of the `t Hooft diagram. We can
remove the $\bar m$ dependence by rescaling
$P_I \rightarrow (2\bar m)^{1\over 2} P_I$ and $\pi^I
\rightarrow 2\bar m \pi^I$ so that the propagator becomes
\eqn\propagator{ \int_0^\infty ds_I
\ \exp\left[ -s_I (P_I^2 +  {\cal W}^\alpha \pi^I_\alpha
+ m)\right].}
This rescaling keeps invariant the measure $d^4 P\ d^2 \pi$
of the zero mode integral.

This piece of the Feynman diagram computation is exactly
what one sees as the spacetime part of the
superstring computation which we reviewed
in the last section. In fact the propagator \propagator\
is the zero slope limit of the open string propagator
evaluated in the Hamiltonian formulation on the worldsheet,
where the Schwinger parameters $s_I$ are coordinates in
scaling regions near the boundaries of the moduli
space of open string worldsheet where open string propagators
become infinitely elongated and worldsheets
collapse to Feynman diagrams. As pointed out in \wittenjones\
and elaborated in more detail in \bcov , integrals over
the moduli space of worldsheets which define topological
string amplitudes localize to these regions. This is how
the topological string amplitude computations discussed in
the last section automatically give results in the field
theory limit. In fact we saw explicitly the $\alpha'$ dependence
cancels out in the final expression of the topological
amplitudes. To make the dictionary complete, the bosonic
and fermionic momenta, $P_{I\alpha\dot \alpha}$ and $\pi_\alpha^I$,
are the zero mode of $i\partial X_{\alpha\dot\alpha}$ and
$p_\alpha$ on the open string propagator. In the exponent
of \propagator, $P_I^2$ is the zero slope limit of the
worldsheet Hamiltonian $L_0 + \bar L_0$, and
and the term $s_I {\cal W}^\alpha \pi_\alpha^I$ comes
from the gluino coupling
$\oint {\cal W}^\alpha p_\alpha$ on the boundary of
the string worldsheet. In this setup,
the superpotential $W(\Phi)$ of the gauge theory
encodes the  information on
the internal Calabi-Yau space.

For an $l$-loop Feynman diagram,
the fermionic momenta $\pi^I$ are parametrized by loop momenta
$\pi^A$ ($A=1,\cdots,l$) as
\eqn\loopmomenta{
 \pi_{\alpha}^I = \sum_{A=1}^{l} L_{IA} \pi_\alpha^A,}
where $L_{IA}=\pm 1$ if the $I$-th propagator is
part of the loop $A$ (taking into account the relative
orientation of $I$ and $A$) and $L_{IA}=0$ otherwise.
Note that, if we view the 't Hooft diagram as the zero
slope limit of the open string worldsheet, we have the
relation
$$ l = 2g + h-1,$$
where $g$ and $h$ are the numbers of handles and
boundaries of the worldsheet. From
the field theory point of view, $h$ is also the
number of 't Hooft index loops.

The computation in \gri\ proceeds by noting that,
in order to absorb the fermion zero modes $\pi_\alpha^A$, we need
to bring down $2l$ gluino fields ${\cal W}_\alpha$. Moreover,
for corrections involving $\epsilon^{\alpha\beta}
\tr \left({\cal W}_\alpha {\cal W}_\beta
\right)$, each 't Hooft index loop can contain at most
two ${\cal W}$ insertions.
Therefore, if ${\cal W}$'s are Grassmannian, it immediately
follows that we need the number $h$ of index loops
is $l+1$ or more in order to absorb the $2l$ fermion
zero modes.\foot{We need $h$ to be $l+1$ or more rather than
$l$ since each propagator is associated to a pair
of index loops going in opposite directions and
a sum over $s_I \pi^I$ along all index loops vanish.}
Since $l = 2g+h-1$, this is possible only when $g=0$,
namely the 't Hooft diagram must be planar.
In this case, the product of
the propagators \propagator\ in the Feynman diagram gives
the factor
\eqn\prodprop{ \prod_{A=1}^l \exp\left( - {\cal W}^\alpha
\sum_I s_I L_{IA}
\pi_\alpha^I \right)
=\prod_{A=1}^l \exp\left( - {\cal W}^\alpha
\sum_{B} M_{AB}(s) \pi^B_\alpha \right)
,}
where $M_{AB}(s)$ is an $l \times l$ matrix defined by
\eqn\defmatrix{ M_{AB}(s) = \sum_I s_I L_{IA} L_{IB}.}
The integration over the fermionic
momenta $\pi^A$ produces the determinant
$ \left( \det M_{AB}(s) \right)^2 $.
This $s$-dependent factor is cancelled out by the integral
over the bosonic momenta $P_I$, which produces
$\left( \det M_{AB}(s) \right)^{-2}$.
Similarly
one can extract the contribution to the $U(1)$ coupling
constants ${\rm Tr} \  {\cal W}_{\alpha}\ {\rm Tr} \  {\cal W}^\alpha$, and
see that they also come only from the planar diagrams.

For non-planar diagrams, we have $l+1-h = 2g > 0$,
and therefore we must have more than two ${\cal W}$'s
on some loop in order to absorb all the fermion zero modes.
This is not possible if ${\cal W}$'s are Grassmannian
variables in the Abelian
configuration relevant for \gri .
Therefore non-planar amplitudes vanish in this case
by the fermion
zero mode integral. This is precisely the part of the
story that is going to change when we consider the
$C$-deformation \cdeform\ of ${\cal W}$.

If ${\cal W}$ is not Grassmannian, we need to take into
account their path-ordering along each index loop when
we take a product of propagators as in \prodprop .
In the last section, we saw that the path-ordered
exponential of ${\cal W}^\alpha p_\alpha$ integrated around
a boundary $\gamma_i$ ($i=1,\cdots,h$) of the worldsheet
gives the factor
\eqn\porderedfactor{ {\cal P} \exp\left( \alpha' \oint_{\gamma_i}
 {\cal W}^\alpha p_\alpha\right) =
\exp\left( -2 \alpha'^2 F^{\alpha\beta}
\oint_{\gamma_i} p_\alpha(\tau) \int^\tau_{o_i} p_\beta \right),}
modulo $\oint_{\gamma_i} p_\alpha = 0$. In the field theory
limit, we regard $\gamma$'s as  't Hooft index loops
and replace
$$ \oint_{\gamma_i} p_\alpha \rightarrow \sum_{I}
 L_{Ii} s_I \pi^I_{\alpha}, $$
where $L_{Ii}$ picks up internal lines $I$'s along the $i$-th index loop
taking into account the relative orientation of $i$ and $I$.
Thus we can write the exponent of \porderedfactor\ as
\eqn\cdeformedfactor{
F^{\alpha\beta} \oint_{\gamma_i} p_\alpha(\tau)
\int^\tau_{o_i} p_\beta \rightarrow F^{\alpha\beta}
\sum_{I>J} s_I L_{Ii} \pi_\alpha^I \cdot s_J L_{Ji} \pi_\beta^J,}
where the inequality $I > J$ is according to the
path-ordering of the edges of the propagators $I,J$
along the $i$-th index loop.

\bigskip\bigskip
\centerline{\epsfxsize 4.8truein \epsfysize 2.7truein\epsfbox{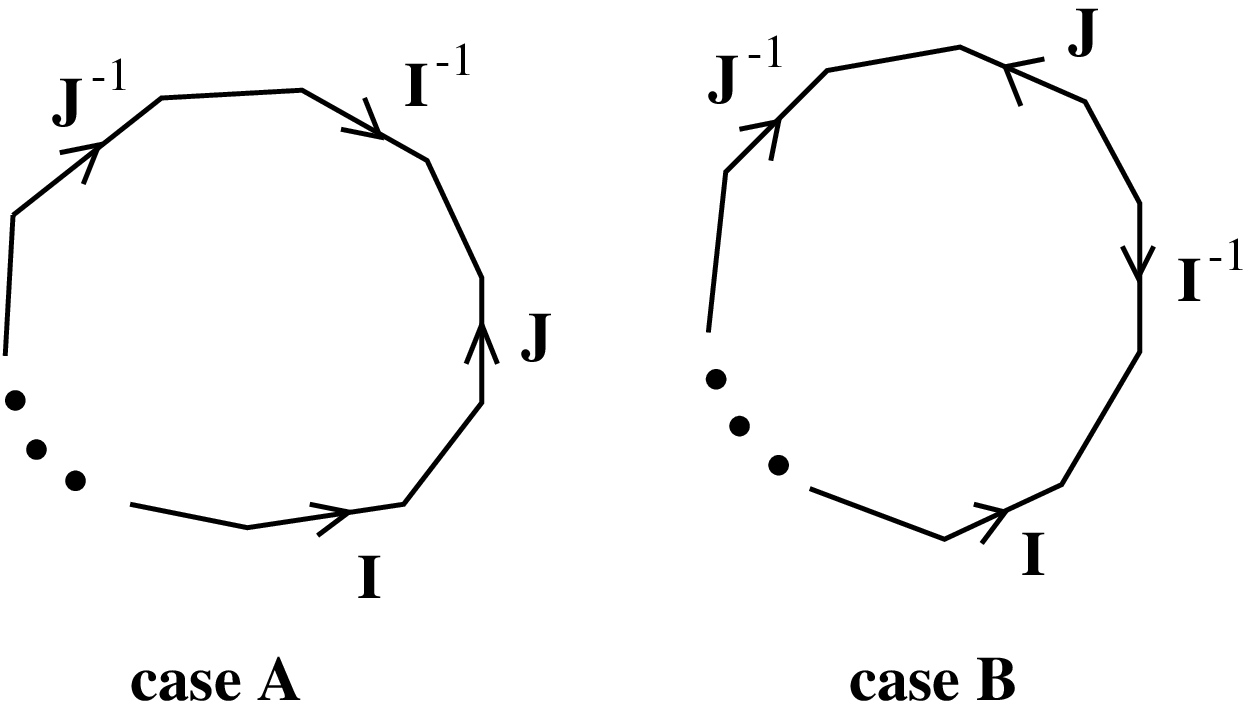}}
\noindent{\ninepoint\sl \baselineskip=8pt {\bf Figure 2}: {\sl
The path-ordered gluino insertion receives
contributions from pairs of edges $I,J$ if they
are oriented as in case $A$. On the other hand,
the contributions cancel in case $B$.}}
\bigskip

In fact the above expression can be evaluated
by a simple set of rules when we have only one boundary
as we will now discuss.
If we have only one boundary every edge appears twice
with opposite orientation (this is in particular consistent
with the fact that $\oint p_\alpha =\sum_I s_I \pi^I L_{Ii}=0$).
Note that the expression \cdeformedfactor\ involves pairs
of distinct edges (for the same edge $F^{\alpha \beta}\pi_\alpha
\pi_{\beta}$ vanishes).  Let us consider two distinct edges
$I$ and $J$ of the boundary (see Figure 2).
The contribution to the exponent of \cdeformedfactor\
vanishes in the case depicted as $B$ in the figure because
the  $\pi^J\pi^I$ terms appear twice with opposite
sign, whereas in the case $A$ they appear twice with the same sign
and so it survives.  Note that in case $A$ if we had the ordering
$IJ^{-1}I^{-1}J$ it would still survive, with an overall
minus sign relative to the case depicted in the figure.
Later we will use this rule to evaluate some examples.

Let us show that
the product of exponential of \cdeformedfactor\ over all index loops,
together with the usual $2(h-1)$ insertions of gluino fields,
absorbs all the fermion zero
modes $\pi_\alpha^A$
and the result of the zero mode integral cancels the $s$-dependent
factor coming from the integral over the bosonic momenta.
We have already seen that this is the case in the topological
string computation in the last section. Here we will show how
this works in the field theory limit. Two gluino fields inserted
on each index loop $\gamma_i$ enforce that the sum over
momenta along the loop vanishes,
\eqn\assumption{ \sum_I L_{Ii} s_I \pi_I = 0. }
Under this condition, we can prove the following identity.
\eqn\intersectionidentity{
\eqalign{ \sum_{i=1}^{h-1}F^{\alpha\beta}
\ \sum_{I>J} s_I L_{Ii}  \pi_\alpha^I \cdot s_J L_{Ji} \pi_\beta^J
&=2 c^{ab} F^{\alpha\beta} \ \sum_I s_I L_{Ia} \pi_\alpha^I\cdot
\sum_J s_J L_{Jb}  \pi_\beta^J\cr
&= 2 c^{ab}F^{\alpha\beta}\ M_{aA}(s) \pi_\alpha^A\cdot  M_{bB}(s)
\pi_\beta^B.}}
To show this, it is most convenient to go back to
the identity \bilinear\ on the string worldsheet
and take the zero slope limit of string theory, where
we can set $p_\alpha = \bar p_\beta$ everywhere on the
string worldsheet. In this limit, \bilinear\ reduces to
$$ -2 F^{\alpha\beta}
\sum_{i=1}^{h-1} \oint_{\gamma_i} p_\alpha(\tau)
\int^\tau_{o_i} p_\beta
= 2 F^{\alpha\beta}
c^{ab}
\oint_a p_\alpha \oint_b p_\beta , $$
since $F^{\alpha\beta}\int p_\alpha \bar p_\beta
= F^{\alpha\beta}\int p_\alpha p_\beta = 0$ by the
symmetry of $F^{\alpha\beta}$ under exchange of
$\alpha,\beta$. The identity \intersectionidentity\
can be obtained by expressing this in terms of the
field theory quantities.\foot{More careful
computation at the leading order in $\alpha'$ shows
\eqn\firstid{
\alpha'^2 F^{\alpha\beta} \int p_\alpha \bar p_\beta=
-{1\over 2}\alpha'^2
c^{ab} F^{\alpha\beta}
\oint_a (p_\alpha-\bar p_\alpha) \oint_b (p_\beta - \bar p_\beta)
 = -2\alpha'^2
 c^{ab}F^{\alpha\beta} \sum_{I,J} L_{Ia}  L_{Jb} \pi_\alpha^I
\pi_\beta^J,}
\eqn\secondid{
-{1\over 2}\alpha'^2
c^{ab} F^{\alpha\beta}
\oint_a (p_\alpha+\bar p_\alpha) \oint_b (p_\beta + \bar p_\beta)
= - 2 \alpha'^2 F^{\alpha\beta}
\sum_{IJ} s_I s_J \cdot L_{Ia}  L_{Jb} \pi_\alpha^I
\pi_\beta^J.}
In the zero slope limit $\alpha'\rightarrow 0$,
the right-hand side of \secondid\ remains finite
if we rescale $s_I \rightarrow
s_I/\alpha'$ (which infinitely elongate open
string propagators) while \firstid\ vanishes in this
limit.
}

Combining the exponential of \intersectionidentity\ with
the $2(h-1)$ insertions of the gluino fields, we find
that the fermion zero mode integral is
given by
\eqn\integrand{\int d^{2l} \pi \
 \prod_{i=1}^{h-1} \epsilon^{\alpha\beta}
M_{iA}(s) \pi_\alpha^A M_{iB} \pi_\alpha^B \
\times \exp\left[
 2 c^{ab} F^{\alpha\beta} M_{aA}(s) \pi_\alpha^A \cdot
M_{bB}(s) \pi_\beta^B\right].}
To evaluate this, it is convenient to make the change
of variables,
$$ \pi_\alpha^A \rightarrow \hat \pi_{\alpha A} =
M_{AB}(s) \pi_\alpha^B.$$
The integral \integrand\ then becomes
$$ \eqalign{ & \left( \det M_{AB}(s) \right)^2
\int d^{2l} \hat \pi_A \
\prod_i \epsilon^{\alpha\beta} \hat\pi_{\alpha i}
\hat \pi_{\beta i} \ \times \exp\left[
2 c^{ab} F^{\alpha\beta} \hat \pi_{\alpha a}
\hat \pi_{\beta b}\right]\cr
&= (\det M_{AB}(s))^2 (F^{\alpha\beta}F_{\alpha\beta})^g.}$$
As in planar diagrams,
the integral over bosonic momenta gives the factor
of $(\det M(s))^{-2}$. So we are left with no
$s$-dependent factor and we just have $(F^2)^g$ (with
some factors of $(2\pi)$ which can be absorbed into the
definition of $F$),
in addition to the $S^{h-1}$ factor.  Moreover we have the
combinatoric factore of $Nh$. The factor of $N$ comes
from the loop with no glueball insertions and the factor
of $h$ comes from the choice of which of the $h$ boundaries
we choose not to put the glueball superfield on.

\subsec{Examples}

It is helpful to illustrate the general
derivation in the field theory limit presented above
by some examples.  Here we will consider three
examples.  In the first example we show how
the computation works for the case of a simple
genus g diagram with one boundary, which in the field
theory computation arises from a $2g$ loop Feynman diagram involving
a single ${\rm Tr} \  \Phi^{4g}$ interaction.  In the second
example we consider a genus 1 diagram with one boundary,
involving two ${\rm Tr} \
\Phi^3$ vertices.  In the third example we consider a diagram
with $g=1$ and $h=2$ involving four ${\rm Tr} \ \Phi^3$ interactions.
The last example is the most interesting one, in the sense
that it involves both the glueball superfield and the $F^2$ term.

\bigskip
\noindent
${\bf  Example~1}$:

{}From a single ${\rm Tr}\ \Phi^{4g}$ vertex we can form
a genus $g$ surface with a single boundary. The Feynman
diagram for this interaction involves $2g$ loops.
 The boundary
consists of $2g$ pairs of oppositely oriented edges each
of which forms one loop.  Along the boundary of the
Riemann surface they are ordered
according to the usual opening up of a genus $g$ surface
in the form
$$I_1I_2 I_1^{-1}I_2^{-1}I_3 I_4 I_3^{-1}I_4^{-1}...I_{2g-1}
I_{2g}I_{2g-1}^{-1}I_{2g}^{-1}$$
According to the rule discussed before, for the path-ordered
gluino insertions we get
$${\rm exp}\Big(\sum_{I=1}^{g}\langle s_{2I-1} \pi_{2I-1} ,
s_{2I}\pi_{2I}\rangle \Big) $$
where $\langle .,.\rangle$ denotes the contraction with
$F^{\alpha \beta}$.  Integration over the fermionic loop momenta
is the same as integration over the ${\cal \pi}_I$ edge
momenta as they are in one to one correspondence.  To absorb
the zero modes we have to bring down each term in the exponent
exactly twice.  This gives the factor
$$(F^2)^g\prod_{I=1}^{2g} s_I^2$$
The bosonic momentum integral (up to factors of $2\pi$
which can be absorbed into the definition of $F$) gives
$1/s_I^2$ for each loop and so the product over all the loops
cancels the $s$ dependence, as expected, leading to $(F^2)^g$.

\bigskip
\noindent
${\bf  Example~2}$:

As our next example we consider a genus 1 diagram with
one boundary formed from two trivalent vertices (see Fig. 3).
The edges along the boundary are ordered as $1\ 2^{-1} \ 3 \
1^{-1}\ 2\ 3^{-1}$.
Thus the path ordered contribution gives
\eqn\rulap{{\rm exp}\Big(- \langle s_1\pi_1 ,s_2\pi_2\rangle  -\langle
s_2 \pi_2,s_3\pi_3\rangle -\langle s_3\pi_3 ,s_1\pi_1\rangle
\Big) }
Note that, as already explained,
 this factor can also be written as the product of
 integral of fermionic
momenta around the two non-trivial cycles of the torus, denoted
by $A$ and $B$ in Fig. 3.  Namely
$$\pi_{{\rm along}\ A}=s_1 \pi_1-s_2\pi_2$$
$$\pi_{{\rm along}\ B}=s_2 \pi_2-s_3\pi_3$$
and we have
$$\langle \pi_{{\rm along}\ A} ,\pi_{{\rm along}\ B}\rangle =
\langle s_1\pi_1 ,s_2\pi_2\rangle  +\langle
s_2 \pi_2,s_3\pi_3\rangle +\langle s_3\pi_3 ,s_1\pi_1\rangle, $$
which is the same as \rulap , up to choice
of orientation of cycles. Here
we have used the fact that $\langle \pi_2 ,\pi_2 \rangle =0$, $etc$.
Writing these in terms of
 the fermionic
loop momenta $\pi_A$ and $\pi_B$ we have
$$\pi_ 1=\pi_A, \ \pi_2=\pi_B-\pi_A, \ \pi_3=-\pi_B$$
which leads to the path ordered contribution
$${\rm exp}\Big[-(s_1s_2+s_2s_3+s_1s_3)\langle \pi_A , \pi_B \rangle
\Big]$$
and integration over the $\pi_A$ and $\pi_B$ leads to the
factor
$$(s_1s_2+s_2s_3+s_3s_1)^2 F^2 $$
The $s$ dependence cancels the bosonic momentum, as can
be readily checked by computation of $(\det M)^2$ where
$$M=\left(\matrix{s_1+s_2&-s_2\cr -s_2&s_2+s_3\cr}\right).$$

\bigskip\bigskip
\centerline{\epsfxsize 4.0truein \epsfysize 2.6truein\epsfbox{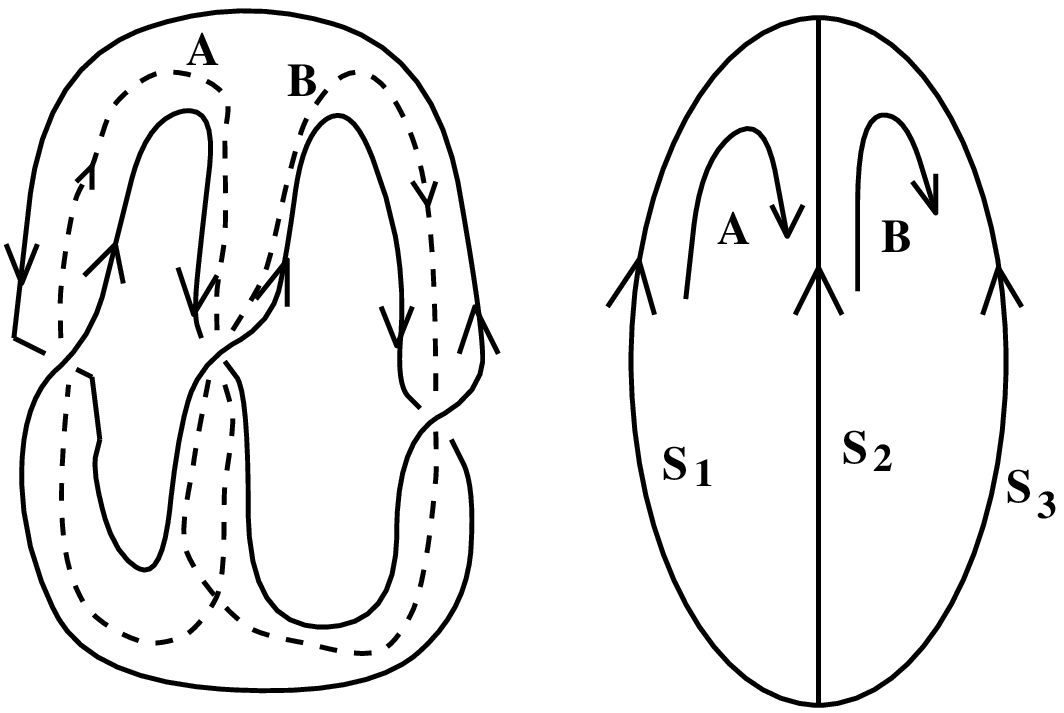}}
\noindent{\ninepoint\sl \baselineskip=8pt {\bf Figure 3}: {\sl
The genus 1 Riemann surface with one boundary, constructed
from two cubic interactions.}}
\bigskip

\bigskip
\noindent
${\bf  Example~3}$:

For a more involved example consider the diagram in a theory
with cubic interactions, drawn in Fig. 4.
  This corresponds
to a diagram with genus 1 and 2 boundaries.  Thus it will
contribute a term $2N (F)^2 \times S$, times the amplitude of the matrix
model, to the superpotential.  The factor $2$ comes from the fact
that we can attach the two ${\cal W}$'s comprising
the glueball field at either of the two
holes, and the factor of $N$ comes from the trace over
the hole where there are no glueball fields.  This diagram has
6 edges with Schwinger parameters $s_I$ with $I=1,...,6$.
The three fermionic {\it loop} momenta we will denote by $\pi_{A,B,C}$.
The two possible choices of the holes for attaching the glueball
field both give the same contribution to the fermionic
momentum integral, namely
$$(s_5(\pi_C-\pi_B)+s_6\pi_C)^2=((s_5+s_6)\pi_C -s_5\pi_B)^2$$
where by square, we mean the $\epsilon^{\alpha \beta}$ contraction.
Similarly the integral over the path ordered integral of ${\cal W}$
can be performed as follows: In this case
only one of the two boundaries contribute because
$s_5\pi_5-s_6\pi_6=0$ (by the absorption of the
fermion zero modes of the glueball
insertion).  The contribution for the larger
boundary is given by the argument we outlined before, as we
order the boundaries according to $1 \ 2^{-1} \ 4 \ 5 \ 3
\ 1^{-1} \ 2 \ 3^{-1} \ 6^{-1} \ 4^{-1}$
and using the fact that $s_5\pi_5=s_6\pi_6$,
by
$${\rm exp}\ \langle s_1\pi_1-s_2\pi_2 ,
s_2\pi_2 -s_3\pi_3 -s_5\pi_5-s_4\pi_4
\rangle .$$
This can also be viewed as the ${\rm exp}
\langle \pi_{{\rm along}\ A}, \pi_{{\rm along}\ B}\rangle $, where
$A,B$ are the two cycles of the torus (see Fig. 4).
Substituting fermionic loop momenta
$$\pi_1 =\pi_A, \ \pi_2=\pi_B-\pi_A, \ \pi_3=-\pi_B$$
$$\pi_4=-\pi_B, \ \pi_5=\pi_C-\pi_B, \ \pi_6=-\pi_C$$
yields
$${\rm exp}\Big( A(s) \langle \pi_A ,\pi_B \rangle +B(s)\langle
\pi_A,\pi_C\rangle +C(s)\langle \pi_B,\pi_C \rangle\Big) $$
 and
$$A(s)=s_1s_2+s_1s_3+s_2s_3+s_1s_5+s_2s_5+s_2s_4+s_1s_4$$
$$B(s)=-s_1s_5-s_2s_5$$
$$C(s)=s_2s_5$$

\bigskip\bigskip
\centerline{\epsfxsize 4.3truein \epsfysize 2.8truein\epsfbox{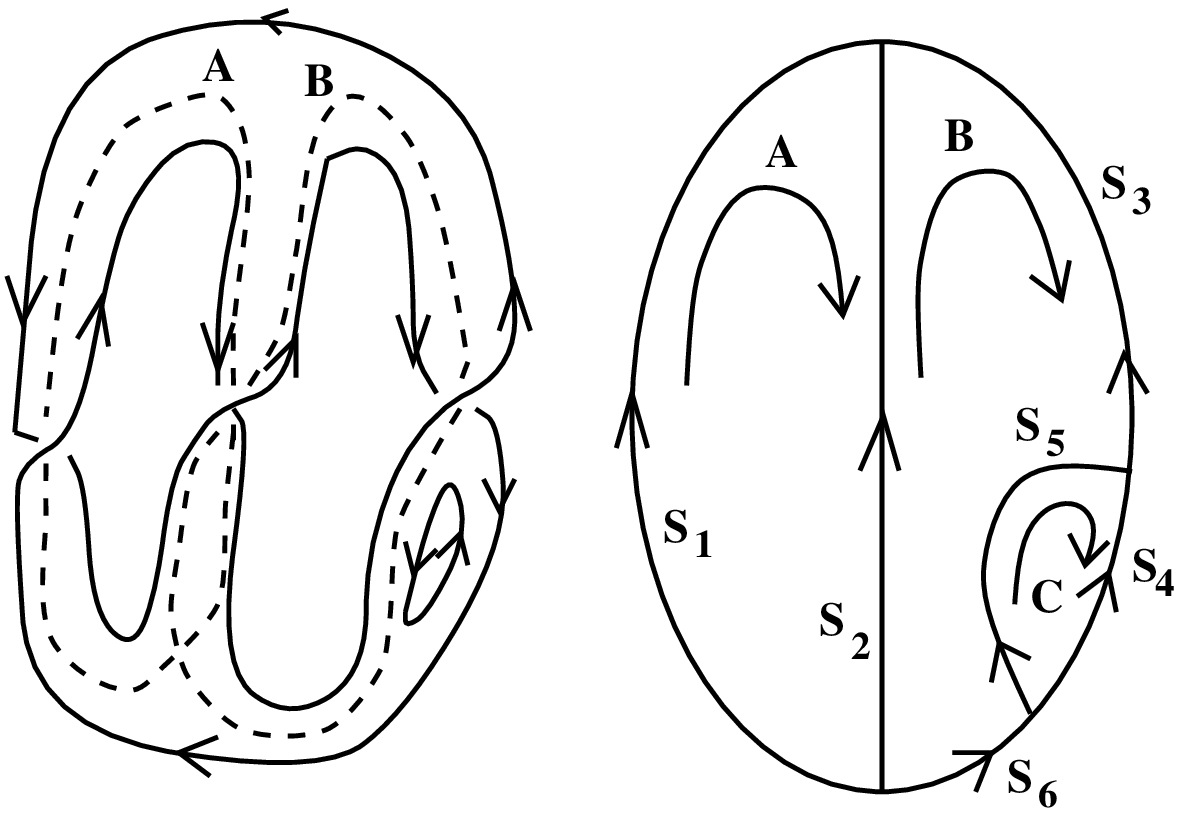}}
\noindent{\ninepoint\sl \baselineskip=8pt {\bf Figure 4}: {\sl
A Riemann surface with genus 1 and two boundaries made out of
four cubic interactions.}}\bigskip

We then need to integrate over the three fermionic loop momenta.
To absorb the $\pi_A$ fermions, we can bring down two $A$ terms
from the exponent, which also absorbs the $\pi_B$ integral and so
we will have to pair it up with the $(s_5+s_6)^2$ in the $S$ contribution
to absorb the $\pi_C$ integral. Or we can bring two $B$ terms which
will have to be paired up with the $s_5^2$ term.  Or we can
bring down one $A$ and one $B$ term which will have to be paired
with the cross term in the $S$ contribution of the form $2(s_5+s_6)s_5$
in the $\pi_B\pi_C$ term.  Putting all these together we find the
$s_i$ dependence is given by $D^2$, where
$$\eqalign{ D & =A(s)(s_5+s_6)+B(s)s_5\cr
&= s_1s_2s_5+s_1s_3s_5+s_2s_3s_5+s_1s_4s_5 \cr
&~~~+s_2s_4s_5+s_1s_2s_6+s_1s_3s_6+
s_2s_3s_6\cr
&~~~+ s_1s_4s_6+s_2s_4s_6+s_1s_5s_6+s_2s_5s_6.}$$
The integration over the bosonic momentum gives the inverse square of
 determinant
of $M$ where $M_{AB}=\sum_I s_I L_{IA} L_{IB} s_I$ is given by
$$M=\left(\matrix{
s_1+s_2&-s_2&0 \cr -s_2 &
s_2+s_3+s_4+s_5&-s_5\cr 0&-s_5&s_5+s_6 \cr} \right) $$
and one easily checks that
$$D=\det  M$$
as expected.

\newsec{Physical interpretation}

We have seen that the connection between matrix model
and ${\cal N}=1$ supersymmetric gauge theories in four dimensions
can be made more canonical, $i.e.$ be extended
 to all higher genus matrix amplitudes if we make the gluino
fields in the path-integral not to be purely Grassmannian.
This lack of anticommutativity breaks Lorentz
invariance, but preserves ${\cal N}=1$ supersymmetry.
In this section we discuss possible physical implications of this
idea.

Even though in this paper we have mainly concentrated on the
chiral sector, corresponding to turning on
$F_{\alpha \beta}$, we could also repeat this
analysis for the anti-chiral sector, by turning on
$F_{{\dot \alpha} {\dot \beta}}$.\foot{In the string
theory context this would lead to a gravitational back-reaction,
which is irrelevant in the field theory limit we are considering.}
  In this Euclidean
context these are independent real numbers, but in the Minkowski
context these are complex
quantities and the reality ($i.e.$, unitarity) conditions dictate
$$F_{\dot \alpha \dot \beta}=(F_{\alpha \beta})^*.$$
This in particular means that for the gluino fields
$\psi_\alpha$ and $\psi_{\dot \alpha}$ in the path-integral
we require,
$$\{ \psi_\alpha, \psi_\beta \} =2F_{\alpha \beta}$$
$$\{ \psi_{\dot \alpha}, \psi_{\dot \beta} \}
=2F_{{\dot \alpha} {\dot \beta}}$$
In the case where $F=0$, $i.e.$ the standard gauge theory context,
we know that the glueball superfield
$$S={1\over 32 \pi^2}\epsilon_{\alpha \beta}
{\rm Tr} \ {\cal W}^\alpha {\cal W}^\beta$$
is the right variable to describe the infrared physics.  Similarly
here, given the link between the full matrix model computation
and gauge theory computation, it suggests that $S$ again
is the right field
in the IR to capture the relevant physics.  We will
assume that to continue to be the case even after
introducing the $C$-deformation.
  In particular in the IR we will
have an effective superpotential
\eqn\supo{W(S)=N {\partial F(S,\lambda ^2 )\over \partial S}+\tau S}
where
\eqn\conm{{\rm exp}\left[
 {1\over \lambda ^2}F(S, \lambda ^2)\right]
={1\over {\rm vol}(U(M))}\int
d\Phi \ {\rm exp}\left[-{1\over \lambda}W(\Phi)\right]}
and $S=\lambda M$ in the above expression.  Moreover
$$ \lambda ^2 =\epsilon^{\alpha \alpha'}\epsilon^{\beta \beta'}
F_{\alpha \beta}F_{\alpha' \beta'} ,$$
in the gauge theory interpretation.
In the corresponding low energy physics we are instructed
to minimize the physical potential
$$V=g^{S{\overline S}}|\partial_S W|^2$$
where $g_{S{\overline S}}=\partial_S\partial_{\overline S} K(S,{\overline
S})$ and $K$ is the (as yet to be computed)  potential coming
from D-term.

 There is a surprise here:
 The IR physics appears to be Lorentz invariant!  Namely
both $S$ and $\lambda^2 =F_{\alpha \beta} F^{\alpha \beta}$ are
Lorentz invariant, and so the $W(S)$ is Lorentz invariant. So there
is no hint in the IR that we are dealing with a theory which
intrinsically breaks Lorentz-invariance.  In other words,
it appears that Lorentz invariance has been restored in the IR!
Even though there are examples where the theory in the IR has
more symmetries than in the UV, for example
theories which have higher dimension operators
violating some symmetry which becomes irrelevant in the IR, it is
amusing that this is appearing also in our case where the fundamental
fields have Lorentz-violating rules for the path-integral.  Note that turning
on $F^2$ does change the expectation value of $S$ at the critical
point and the critical value of $W$, but in a Lorentz-invariant way.
It is tempting to speculate about the potential realization of this
idea in Nature.  In particular this would be consistent
with the macroscopic existence of Lorentz invariance, which
could get violated
at higher energies.  This is even more tempting since from
the viewpoint of the relation of ${\cal N}=1$ supersymmetric
gauge theories and matrix model, the $C$-deformation is forced on us!
It would be interesting to explore the signature of the $C$-deformation
for potential observations in the accelerator physics or cosmology.

\subsec{Pure ${\cal N}=1$ supersymmetric Yang-Mills revisited}

\lref\distler{
J.~Distler and C.~Vafa,
``A critical matrix model at $c = 1$,''
Mod.\ Phys.\ Lett.\ A {\bf 6}, 259 (1991).
}

\lref\penner{R.~C.~Penner, ``Perturbative series and the
moduli space of Riemann surfaces,'' J. Diff. Geom. {\bf 27}
(1988) 35.}

Let us consider the special case of pure ${\cal N}=1$ Yang-Mills,
deformed by turning on $F_{\alpha \beta}$.  This will be a leading
piece of the superpotential
of many other theories, in the limit where $S$ is small and so higher
powers of $S$ can be ignored
in the glueball superpotential.  In this case, the 
partition function of the matrix model  
\conm\ is entirely
given by the measure factor, ${\rm log} \ {\rm vol}(U(M))$, 
which has been shown \oova\
to give the partition function of $c=1$ at self-dual radius \distler.
We have (up to an addition
of an irrelevant constant ${1\over 12} \lambda^2 {\rm log}\lambda$)
$$F(S, \lambda^2)={1\over 2}S^2 {\rm log}S
-{1\over 12}\lambda^2{\rm log} (S/\lambda ) +
\sum_{g>1} {B_{2g}\over 2g(2g-2)}\cdot {\lambda^{2g}\over S^{2g-2}}$$
This can be written in a more unified form, up to an addition
of  ${-1\over 2} S^2 {\rm log \lambda}$ which in the expression
for the superpotential can be absorbed into redefining the
coupling constant $\tau$,
$${1\over \lambda^{2}}F(S, \lambda^2)=
{1\over 2}{S^2\over \lambda^2} {\rm log}\left({S\over
\lambda}\right)
-{1\over 12}{\rm log}\left( {S\over \lambda}\right) +
\sum_{g>1} {B_{2g}\over 2g(2g-2)}\cdot \left(
{\lambda \over S}\right)^{2g-2}.$$
It is natural to define a rescaled dimensionless
 glueball field $\mu =S/\lambda$.  In
terms of this we have
$${1\over \lambda^2}F={1\over 2}{\mu^2} {\rm log}{\mu}
-{1\over 12}{\rm log} {\mu } +
\sum_{g>1} {B_{2g}\over 2g(2g-2)}\cdot \mu^{2-2g}$$
This leads to the superpotential
\eqn\vyg{{1\over \lambda} W=N
\left( {\mu {\rm log}\mu}-\sum_{g>0}{B_{2g}\over 2g}
\mu^{1-2g}\right)
+\tau \mu}
Note that this $\tau$ defers from the bare $\tau_0$ in the
gauge theory by
$$\tau=\tau_0+N{\rm log }\lambda/\Lambda_0^3$$
 where
$\Lambda_0$ is the cutoff where the bare coupling $\tau_0$ is defined.
Note that writing the superpotential in term of the new $\tau$, undoes
the dimensional transmutation.  In other words, we now have gotten rid
of $\lambda$ and recaptured it in term of the coupling constant $\tau$
which does not run.  Put differently, $\tau$ denotes the coupling
constant of the gauge theory at the scale set by $\lambda$.
It is interesting to note that $\vyg$ is the generating function
for the Euler character of moduli space of Riemann surfaces with
one puncture
\refs{\penner,\distler}.

Note that the superpotential \vyg\ is a generalization of the
Veneziano-Yankielowicz superpotential \ref\vy{
G.~Veneziano and S.~Yankielowicz,
``An effective Lagrangian for the pure ${\cal N}=1$
supersymmetric Yang-Mills theory,''
Phys.\ Lett.\ B {\bf 113}, 231 (1982).
},
taking the $C$-deformation
into account.  The fact that many different powers of $\mu$ enter
is because $F$ `carries' an $R$ charge and with respect to that
$S/\lambda$ is neutral and so in principle arbitrary powers
of it can appear.  Here we are predicting in addition  very
definite coefficients for these terms. We expect, as in the case
of the Veneziano-Yankielowicz potential,
the measure of the gauge theory should somehow
dictate this structure, but we do not, at the present, have
a direct gauge theory derivation of this.

Let us analyze the critical points of this superpotential.
We need to solve $dW=0$ (again we reabsorb a constant
term in the shift of $\tau$):
$${1\over N\lambda}{dW\over
d\mu}={\rm log} \mu+\sum_{g>0}{(2g-1)\over 2g}B_{2g} \mu^{-2g}+
{\tau\over N}=0$$
(It is amusing to note that $dW/d\mu$ is the partition function
of the Euler character of the moduli space of doubly punctured Riemann
surfaces, up to addition of $\tau/N$.)
If $\tau \ll 0$ $i.e.$ if $F$ is much smaller than the
physical scale of the original gauge theory, then we have
$$\mu \sim e^{-\tau/N}$$
This in particular is consistent with dropping the
terms with negative powers of $\mu$ in \vyg\ because
$\mu \gg 1$ (this is self-consistent, $i.e.$ $\langle S\rangle \gg F$).
As we increase $\tau$ the correction terms to VY potential
become more relevant.  Let us define $q=e^{\tau/N}$  Then
we expect, after minimization, to have an expansion
$$W(\mu)\big|_{{\rm min}}=N\lambda\left(
q^{-1}+\sum_{n\geq 1} a_n q^{2n-1}\right)$$
for some computable $a_n$.  It would be interesting
to see if this function has any interesting modular
properties.

\subsec{Non-perturbative completion of $W$ and baryons}

{}From the definition of $F(S,\lambda^2)$ \conm\ as the full
free energy of the matrix model, it is clear that we cannot
stop to all orders in perturbation theory, and in particular
we have to have a full definition of the matrix integral, including
its non-perturbative completion.  This is unlike \dv\ where we
could restrict attention simply to planar diagrams of the matrix
model defined by Feynman perturbation theory.  Thus it is natural
to ask how do we give the full non-perturbative definition of $F$
or the associated physical superpotential \supo .

In order to get insight into this question it is useful to
trace back, within string theory, what turning on the graviphoton
does. On the dual gravity side, the theory is an ${\cal N}=2$ supersymmetric
theory broken down to ${\cal N}=1$ by some flux.
The topological string amplitudes do not depend on the choice
of flux \vaug .  Thus we can ask how does turning on graviphoton
field strength affect the ${\cal N}=2$ theory.

This question was studied at length in \ref\govam{R.~Gopakumar and C.~Vafa,
``M-theory and topological strings. I,''
{\tt arXiv:hep-th/9809187};
``M-theory and topological strings. II,''
{\tt arXiv:hep-th/9812127}.
}.  The main idea is to relate
the turning on of the graviphoton field strength, as giving
rise to a correction to $R^2$ terms which are captured by
Schwinger like computation.  Recall that graviphoton couples
to D-branes with charge proportional to their BPS mass.
 Motivated
by this correspondence it is natural to write the superpotential $W$
as coming from such a computation.  For example for the case
of pure Yang-Mills we would be led to
\eqn\tsu{{1\over \lambda} W(\mu)=N\int_\epsilon^\infty {ds\over s^2}
\left( {s/2\over {\rm sinh}(s/2)}\right)^2 e^{-s \mu}}

This suggests a non-perturbative completion of the superpotential
relevant for the cases with smaller values of $\mu$, including
terms of the form $\sim e^{2n\pi i\mu}$, as is familiar from the 
Schwinger computation.  If ${\rm Im} \mu ={\rm Im}(S/|F|)\gg 1$
these effects are
small. Note that for pure Yang-Mills, these corrections,
on the matrix model side, would be invisible:  Since $\mu =S/F=
\lambda_s M/\lambda_s=M$ these terms correspond to ${\rm exp }(2\pi i M)$
which is $1$ and do not depend on $M$.  Thus the ambiguity
of the map between the matrix model and gauge theory data
allow for such additions.

\lref\wittenbaryon{
E.~Witten,
``Baryons and branes in anti de Sitter space,''
JHEP {\bf 9807}, 006 (1998);
{\tt arXiv:hep-th/9805112}.
}

\lref\gross{
D.~J.~Gross and H.~Ooguri,
``Aspects of large $N$ gauge theory dynamics as seen by string theory,''
Phys.\ Rev.\ D {\bf 58}, 106002 (1998);
{\tt arXiv:hep-th/9805129}.
}

Recall that in the string theory
realization the wrapped brane corresponds to
baryons
\refs{\wittenbaryon, \gross}
as the corresponding brane is pierced by $N$ units of
RR flux. Even though in the ${\cal N}=1$ theory
these are not as part of the excitations of the theory
(as one would have to supply the quarks as probes)
nevertheless it is striking that they can be used
to reproduce the superpotential.
Therefore it is natural to interpret
the full superpotential $W$ \tsu\ as obtainable from the
baryon/anti-baryon pair production effect. It would
be interesting to better understand this statement
from the field theory side.
 It is amusing
to note that this includes the Veneziano-Yankielowicz part
of the superpotential as well, suggesting a new interpretation
for it.  For more general ${\cal N}=1$ theories one would expect
that there would be similar completions of the perturbative
computation, similar to that studied in \gova\ in the context
of A-model topological strings.

\bigskip
\bigskip

\centerline{{\bf Acknowledgments }}

\bigskip
We are grateful to N. Berkovits for valuable discussion
about the covariant quantization of superstring.  We would also
like to thank R. Dijkgraaf, M. Grisaru, S. Minwalla, Y. Okawa, J. Schwarz,
P. van Nieuwenhuizen, N. Warner, E. Witten, and S.-T. Yau 
for useful discussions.

H.O. thanks the theory group at Harvard University for
the hospitality. C.V. thanks the hospitality of the theory group at
Caltech, where he is a Gordon Moore Distinguished Scholar.

The research of H.O. was supported in part by
DOE grant DE-FG03-92-ER40701.  The research of C.V. was supported
in part by NSF grants PHY-9802709 and DMS-0074329.

\listrefs

\end